\documentclass[fleqn, 12pt]{article}
\usepackage{graphicx}
\usepackage{natbib}

\usepackage{xcolor}

\begin{document}


\begin{center}

{\bf{ELECTROSTATIC LOFTING OF DUST GRAINS FROM THE SURFACES OF THEBE AND AMALTHEA}}

{\it\bf N. Borisov$^{1,2}$, H. Kr\"uger}$^{2}$\\
{$^{1}$IZMIRAN, 108840, Moscow, Troitsk, Russia\\ $^{2}$MPI for Solar System Research, 37077, G\"ottingen, Germany}


\end{center}
\begin{abstract}
\noindent Energetic electrons from the inner radiation belt provide
significant electric charging of the surfaces of Jupiter's moons Thebe and Amalthea
whose orbits are located within this radiation belt. We estimate theoretically the electric fields
in the vicinity of the polar regions of Thebe and Amalthea and argue that
these fields are sufficient for lofting of micron and submicron-sized dust grains from the surfaces
of the moons. Thus, the lofting of charged dust grains in the electric fields can be considered as an additional
source supplying dust to the gossamer rings in addition to dust ejection by
micrometeoroid impacts onto the moons' surfaces.
The suggested mechanism can explain qualitatively some peculiarities of the
dust grain distributions within the gossamer rings.
\end{abstract}


\section{Introduction}
Jupiter has a highly structured dust ring system that extends along the planet's equatorial plane.
It consists of the main ring, the halo, two
gossamer rings and the Thebe extension \citep{burns1984a,showalter1987,burns2004}.
This dust system  was investigated 
by the space missions Voyager 1/2, Galileo, Cassini and New Horizons
\citep{smith1979a,owen1979,ockert-bell1999,porco2003,burns2004,showalter2007,throop2016} and also by
telescopes in space (Hubble) and from the Earth (Keck) \citep{depater1999,depater2008,showalter2008}.

The main ring is located between $1.72 R_J$ and $1.806 R_J$ from Jupiter, where $\mathrm{R_J}$ is the radius of Jupiter
 $\mathrm{R_J = 71,492~km}$. This ring has 
an optical depth $\tau \geq 10^{-6}$ and a thickness perpendicular to
Jupiter's equatorial plane of approximately $30 - 100$ km.
Its radial extension
is of the order of 6000~km. Interior to the main ring
 the halo  has a rather wide $(\sim 2 \cdot 10^4$~km) and thick $(\sim 10^4$~km) structure with
an optical depth of $\tau\sim 10^{-6}$, similar to that of the main ring.
Jupiter's gossamer ring system was detected out to $3.5\,\mathrm{R_J}$ from the planet
 on optical images and further beyond with
in-situ detections, see, e.g. \citet{showalter2008}. Two small moons, Amalthea and Thebe, orbiting Jupiter
within the ring region (at $2.5\mathrm{R_J}$ and $3.1\mathrm{R_J}$ distance from the planet),
are generally considered as the major dust sources
feeding the gossamer rings via the ejection of particles released due to
bombardment of the surfaces of these moons by micrometeoroids.

The Amalthea ring is brighter and narrower
than the Thebe ring. It is situated between the main ring and the orbit of the moon
Amalthea. This ring has a slightly triangular shape
when viewed edge-on and its optical depth is
$\tau\sim 10^{-7}$. The dust particles detected in the ring have sizes of $0.5 - 2.5$ $\mathrm{\mu m}$
\citep{krueger2009b}.
 The Thebe ring is broader and fainter than the Amalthea ring, with an optical depth $\tau\sim 3\cdot10^{-8}$.
 It terminates at the
orbit of Thebe. The sizes of dust particles in this ring are similar to those measured within the Amalthea ring.
Outside Thebe's orbit there is one more (even fainter)
structure which is known as the Thebe extension. The distribution of dust in the
gossamer rings has some peculiarities: The thickness of these rings
is approximately determined  by vertical excursions of the corresponding
moons during their orbital motion around Jupiter, and the ring's upper and lower edges are much brighter
than their central
parts. In both rings the height integrated concentration of dust grains has its maximum
just interior to the corresponding moon's orbit.

It is commonly believed that dust forming the gossamer rings is supplied
exclusively due to continuous bombardment by micrometeoroids onto the surfaces of Thebe and Amalthea
\citep{krivov2002a,dikarev2006}. On the other hand, electric charging of dust in planetary rings has also
been considered \citep{graps2008}.
It is usually accepted that dust grains lofting from
the surfaces of cosmic objects without an ionosphere (e.g., the Moon and asteroids) is caused
by the electric forces due to surface charging of dust grains located on the surface, see,
e.g. \citet{lee1996,horanyi1998b,colwell2005}. The dust surface charge is provided by photoelectron emission
caused by solar UV radiation and plasma impacting the grain surface.
Such a mechanism can be  efficient if electric forces acting on dust grains
near the surface are strong enough to overcome  gravity and adhesion
\citep{hartzell2011,hartzell2013a,hartzell2013b,kimura2014,wang2016}.
On the surfaces of
objects without an ionosphere, strong
electric fields can be formed on uneven surfaces near the
terminator and at the un-illuminated side because the fluxes of solar wind electrons and protons
that hit the surface are not the same \citep{borisov2006}.

Jupiter
has a strong magnetosphere which prevents the solar wind plasma from penetrating into it.
The plasma composition of Jupiter's inner magnetosphere is complicated.
The plasma in Jupiter's inner magnetosphere was modelled based on the
Pioneer and Voyager data provided by \citet{divine1983}. Later on, this model was improved with results
from more recent space missions \citep{garrett2005}.
Unfortunately, up to now only a few plasma observations are available for the
inner magnetosphere of Jupiter.
As a consequence,
exact plasma parameters in this region are still not well-known, however,
the Juno mission \citep{bolton2010} that has recently arrived at Jupiter
 will help to solve this problem in the future.
Different plasma populations were identified in Jupiter's inner magnetosphere. They are
variable in
time, so that
existing models can represent only an averaged distribution. First, it is thermal plasma
approximated by the Boltzmann distribution
with some finite temperature. Second, energetic
electrons and protons that form the radiation belts exist in the inner
magnetosphere. Up to now only the fluxes of electrons with  energies  $W \geq 160$ keV
and protons with  energies above $W\approx 1.2$ MeV were measured in the inner radiation belt,
\citep[see, e.g.,][]{khurana2004}.

The fluxes of energetic electrons are very large. Depending on the
model, in the vicinity of Thebe or Amalthea, they
can be of the same order of, or even exceed, the fluxes of thermal electrons.
The probability that such energetic electrons impact micron-sized or smaller dust particles in the gossamer rings
is negligible, hence charging of small dust particles due to energetic electrons with energies
exceeding $W>100$ keV can be neglected. At the same time such energetic electrons contribute to the charging of the
moons' surfaces, leading to the formation of strong surface electric fields.

Up to now the role of  surface charging in dust grains lofting from Thebe and Amalthea has not been
investigated. The aim of the present paper is to investigate theoretically the electric fields near the
surfaces of these two jovian moons (in particular in their polar regions)
due to the charging by the fluxes of electrons and protons. 
 As the magnetized electrons and protons propagate along the magnetic field lines, their fluxes to the unit square
on the surface are proportional to
$\cos\theta$ (where $\theta$ is the angle between the magnetic field and the
normal to the surface). Such fluxes onto the surface reach their maximum values in the vicinity of the polar
regions where the angle $\theta$ is small. Hence, the strongest negative potential on the surface
is expected in the polar regions. This provides more favourable conditions for charged particles to leave the moons.
Simultaneously, the solar UV radiation acts
only on the sunlit side and its flux is practically tangential to the
surface in the polar regions. So, it does not produce any significant charging there.
Below we estimate the sizes of dust grains
that can be lofted from the surfaces by the predicted electric fields.

\section{Basic equations}

\label{sec_2}

First we need to introduce the distribution of plasma components in the inner radiation belt of Jupiter. As we
concentrate on the surface charging and dust grain lofting from Thebe and Amalthea we are interested in the plasma
distribution at the radial distances $\approx 3.1\,\mathrm{R_J}$ (orbit of Thebe) and $\approx 2.5\,\mathrm{R_J}$ (orbit of Amalthea).
As mentioned above, parameters of the inner magnetosphere of Jupiter are still not well-known.
Nevertheless, based on the existing models we introduce the Boltzmann
distribution  which describes thermal electrons:
\begin{equation}
F_e^{(th)}=\frac{N_{e,th}}{(\sqrt{\pi}V_{e,th})^3}\exp\left (-\frac{v^2}{V_{e,th}^2}\right ) \label{eq_1}
\end{equation}
and thermal ions
\begin{equation}
F_i^{(th)}=\frac{N_{i,th}}{(\sqrt{\pi}V_{i,th})^3}\exp\left (-\frac{v^2}{V_{i,th}^2}\right ). \label{eq_2}
\end{equation}
Here $N_{e,th}$, $N_{i,th}$ are the concentrations of electrons
and ions, and $V_{e,th}$, $V_{i,th}$ are their thermal velocities.
 The temperatures are assumed to be equal to each other $T_{e,th}=T_{i,th}=T_{th}$.
 Note that in the inner magnetosphere the most abundant ions
are oxygen $O^{+}$ $(\sim 20\% )$ and sulphur $S^{+}$ ($\sim 70\%)$ instead of protons.

The distribution of energetic electrons and protons in the radiation belt can
be modeled as a power-law distribution (it is often called a kappa distribution) \citep{divine1983}:
\begin{equation}
F_{\alpha}(W)=N_{\alpha,h}\left (\frac{m_{\alpha}}{2\pi W_{0,\alpha}}\right )^{3/2}\frac{\Gamma (\kappa +1)}
{\kappa^{3/2}\Gamma (\kappa -1/2)(1+W/\kappa W_{0,\alpha})^{\kappa +1}}. \label{eq_3}
\end{equation}
Here, $N_{\alpha,h}$ with $\alpha =e,i$ are the
concentrations of hot electrons and protons in the radiation belt,
$W_{0,\alpha}$ is the characteristic energy. Parameters $W_{0,\alpha}$ and
$\kappa$ should be chosen such that they represent more or less accurately the
energy distributions of electrons and protons experimentally measured in
the radiation belt.

Our aim is to provide only estimates  of dust charging. So,
we shall use rather simple power-law distributions that make it possible to obtain the results
analytically. Different electron distributions exist
in the radiation belts. Among them are the so-called loss-cone distribution and
the pancake distribution. The loss-cone distribution
corresponds to the case when for a given energy, $W$, in the equatorial plane the transverse
 velocity of electrons is much smaller than the longitudinal velocity.
Indeed, energetic electrons in the radiation belt
bounce between the reflection points where the magnetic field reaches such high values that
the  longitudinal velocity becomes equal to zero. As in the magnetic
field for strongly magnetized electrons the magnetic moment $\mu=v_{\perp}^2/2H$ is conserved,
the longitudinal velocity is presented in the form $v_z^2=2(W/m-\mu H(z))$. It
follows from this relation that for a given energy the longitudinal velocity
decreases while electrons move along the magnetic field line towards the
poles. At the same time the longitudinal velocity reaches its maximum value
near the equator.
 In the theoretical analysis for such a distribution of electrons the following model is often used:
\begin{equation}
F_e^{(h)}=\frac{24}{\pi^2}N_{e,h}\frac{v_{\perp}^2V_{e,h}^3}{(v_z^2+v_{\perp}^2+V_{e,h}^2)^4}.
\label{eq_4}
\end{equation}
Here $N_{e,h}$ is the concentration of energetic electrons, $v_z, v_{\perp}$
are their velocities along and across the magnetic field line, and
$V_{e,h}$ is the characteristic velocity of energetic electrons.  If we present
$v_{\perp}=v\sin\alpha$ (where $\alpha$ is a pitch angle) it becomes clear that
the distribution (\ref{eq_4}) has a maximum for large pitch angles $|\alpha |\approx \pi/2$. The
corresponding concentration $N_{e,h}$ we present in  the form $N_{e,h}=\eta_e
N_{e,th}$, where $\eta_e <<1$ is a relative concentration of energetic electrons
with respect to thermal electrons. Note that in the model by \citet{garrett2005} a very strong
dependence on the pitch angle $\propto \sin^{40}\alpha$ was introduced in
addition to the isotropic distribution \citep{levin2001}. In our rather crude estimates we
shall use the distribution (\ref{eq_4}) which qualitatively describes the anisotropy of
the electron distribution.

Another possible distribution of electrons is a pancake distribution. According to
the analysis of the synchrotron emission in the inner radiation belt of Jupiter
at small L-shells ($L<2$) the
distribution of energetic electrons near the equator has some fraction which can be described as a pancake
distribution, see, e.g. \citet{garrett2005}. The peculiarity of this distribution
is that the  maximum corresponds to small pitch angles. Such a distribution can be
modelled analytically as the following:
\begin{equation}
F_e^{(h)}=\frac{48}{\pi^2}N_{e,h}\frac{v_z^2V_{e,h}^3}{(v_z^2+v_{\perp}^2+V_{e,h}^2)^4}.
\label{eq_5}
\end{equation}
For $v_z=v\cos\alpha$  the distribution ({\ref{eq_5}) indeed has its maximum at small pitch
angles. As the moons Amalthea and Thebe orbit Jupiter at larger L-shells,  such a
pancake distribution will be absent in our calculations.

For energetic protons in the radiation belt we introduce the distribution:
\begin{equation}
F_i^{(h)}=\frac{16\eta_iN_0}{\pi^2}\frac{v^2 V_{i,h}^3}{(v^2+V_{i,h}^2)^4}. \label{eq_6}
\end{equation}
Contrary to  eq.~(\ref{eq_3}) our model~(\ref{eq_6}) describes qualitatively the decrease of the proton distribution not only
for high energies $v^2 >>V^2_{i,h}$ but also for small energies $v^2
<<V^2_{i,h}$.  It looks reasonable because the distribution of energetic
protons should decrease for small enough energies.
To preserve the quasineutrality of the plasma we assume that
$N_{e,th}=N_0$, $N_{i,th}=(1+\eta_e -\eta_i ) N_0$. Parameters $N_0$, $\eta_e$,
$\eta_i$, $v_{e,h}$, $v_{i,h}$ are to be determined from the experimental data.

We need to take into account the emission of secondary electrons from the
moon's surface  due to bombardement by fast electrons because secondary electrons
can make a significant contribution to
surface charging \citep{whipple1981}. The amount of newly created secondary electrons at the surface is
determined by the energy distribution of primary electrons hitting the surface  and the
yield $\delta (W)$:
\begin{equation}
\delta (W)\approx 7.4 \delta_m \frac{W}{W_m}\exp\left [-2\left (\frac{W}{W_m}\right )^{1/2}\right ],
\label{eq_7}
\end{equation}
 where $W=mv^2/2$ is the energy of a primary electron, $\delta_m$ is a maximum yield, and $W_m$
 is the energy that corresponds to the maximum yield \citep{whipple1981,horanyi1993c}. Parameters $\delta_m$, $W_m$
 depend on the material of the surface. For a
thermal distribution of primary electrons (see, eq.~\ref{eq_1}) the amount of secondary
electrons in the range of energies $W, W+dW$ is
\begin{equation}
\frac{dN^{(th)}_{e,sec}}{dW}=\frac{2}{\sqrt\pi}\delta (W)N_{e,th}
\exp\left (-\frac{W}{W_{e,th}}\right )\frac{\sqrt W}{W_{e,th}^{3/2}}.
\label{eq_8}
\end{equation}
Here $W_{e,th}=mV_{e,th}^2/2$ is the thermal energy. Similarly, we present the
production of secondary electrons by energetic primary electrons (eq.~\ref{eq_4}):
\begin{equation}
\frac{dN^{(h)}_{e,sec}}{dW}=8\delta (W)N_{e,h}W_{e,h}^{3/2}\frac{W_{\perp}{\sqrt W}}{(W+W_{e,h})^4},
\label{eq_9}
\end{equation}
where $W_{e,h}=mV_{e,h}^2/2$. The energy distribution of the secondary
electrons is usually   modelled as a Boltzmann distribution with a low
temperature $T_s \sim 2\,\mathrm{eV}$. Later on, for simplicity, we assume that all secondary electrons
are born with the same velocity $v_s =(2 T_s/m)^{1/2}$.  The equations presented above
 make it possible to estimate the stationary electric potential near
the surface of the moon. Only a yet unknown electric potential $\varphi$ should
be introduced in the distribution functions of electrons and ions near the surfaces of the moons.

\section{Calculation of the electric potential on the surface of a large body in a magnetized plasma}

\label{sec_3}

We would like to estimate the electric potential on the surface of a
given moon in the vicinity of its polar region. Here, we assume
Jupiter's magnetic field lines to be orthogonal to the surface. The Larmour radii
of  (not only thermal but
also energetic) ions are much smaller than the sizes of Thebe and Amalthea.
Therefore, it is possible to define two planes.
The first one is Jupiter's equatorial plane, and the orbital planes of both moons are very
close to this plane. The second plane corresponds to the
 magnetic equator of Jupiter. The angle between these two planes is $\beta \approx 10^{\circ}$.
Suppose that at the magnetic equator far from the moon
(where the electric potential tends to zero, i.e. $\varphi \approx 0$)
the distribution functions of charged particles are given by equations presented in Section~\ref{sec_2}.
Note that we use non-relativistic expressions for the magnetic moment and the energy of electrons.
Electrons become relativistic if their kinetic energies exceed $W\approx mc^2\approx 0.5\,\mathrm{MeV}$.
For smaller energies electrons  can be considered as non-relativistic. But even for electrons with
 energies $W\sim 1$ MeV relativistic corrections are rather moderate and
can be neglected for our estimates.

The magnetic field $H$ for a given L-shell changes  along the magnetic field line.
We therefore use the invariant parameters $\mu$
and $w=v_{\perp}^2+v_{z}^2$ instead of
velocities $v_{\perp}$ and $v_{\|}$.
The element of phase space in these parameters takes the form
\begin{equation}
2\pi v_{\perp}dv_{\perp}dv_{z}=\pi\frac{dw H d\mu}{\sqrt{w-H\mu}} \label{eq_10}.
\end{equation}
As the magnetic field grows towards the poles, the distribution of charged particles
(first of all electrons) along the magnetic field line changes. Indeed,
electrons with small longitudinal velocities are reflected where $w=2\mu H$. But in our case
this effect is small. Indeed, in the dipole magnetic field for a given L-shell and
the
moon's orbital inclination with respect to the magnetic equator of $\beta = 10^{\circ}$,
the growth
of the magnetic field is only $\approx 12\%$ with respect to the value at the equator.
This means that electrons with a pitch angle at the magnetic equator
$\alpha \leq 70^{\circ}$ are able to reach a given moon (Thebe or Amalthea). But these electrons
deposit their charge on the moon's surface only if the longitudinal energy at
the magnetic equator $mv_{z}^2/2$ exceeds the potential energy of the surface $|e\varphi_s|$.
Taking into account that our estimates are rather simple (first, because
our present knowledge of the electron and ion distributions
is not sufficient and second, the real magnetic field deviates from a dipole field)
small variations of the magnetic field strength along the magnetic field line
at the angles $\beta \leq 10^{\circ}$ can be neglected.

Near the surface of the moon the electric potential $\varphi$ tends to equalize the
fluxes of electrons and ions hitting the surface. Due to this, the
variable $w$ in the distribution of electrons  changes to $w-2e\varphi /m$.
In this paper we present
rather simple estimates of the surface charging and assume that the electric and magnetic fields
vary only in one direction (along the magnetic field line).
Therefore, we introduce the system of coordinates in which the
z-axis is directed along the magnetic field, and the x-axis is in the radial direction.
With the help of the model distributions presented in Section~\ref{sec_2} and taking into account that
the electric field varies only along the z-axis we can calculate
the electric charge that reaches the surface on a
unit square per second.
Note that the angular velocity with
which  the magnetosphere of Jupiter rotates deviates from the orbital angular velocities
of the moons Thebe and Amalthea.
This leads to the radial electric field, the so-called v $\times$ B field.
According to our estimates,  this field is much smaller than the polarization electric field
caused by the charging of the surface and to a first order approximation it does not influence
 the lofting of dust grains.
At the same time this electric field determines the dynamics of dust
particles on rather long time scales after they leave the vicinity of the
surface.

Under stationary conditions the electric charge that  electrons carry to the
unit square on the surface
should be equal to the electric charge that is carried by ions.
From this relation the electric potential on the surface $\varphi_s$ is obtained. As the
non-disturbed flux of electrons is larger than the flux of ions,
the electric potential $\varphi_s$ should be negative.
This means that electrons are retarded while  ions are accelerated towards the surface.

We shall estimate the fluxes of electrons and ions that deposit electric
charges on the surface of the moons at their polar regions where the
z-axis is assumed to be normal to the surface.
The flux of thermal electrons along the z-axis
is
\begin{equation}
P_{e,z}^{(th)}=\frac{N_0 V_{e,th}}{2\sqrt\pi}\exp\left (\frac{e\varphi_s}{T_{th}}\right ). \label{eq_11}
\end{equation}
Similarly, the flux of thermal ions is
\begin{equation}
P_{i,z}^{(th)}=\frac{(1+\eta_e-\eta_i )N_0V_{i,th}}{2\sqrt\pi}. \label{eq_12}
\end{equation}
Note that this flux does not depend on the electric potential $\varphi$. The
reason for this is that we consider the 1-D case assuming that the electric field
is directed only along the z-axis. In this model the flux of propagating
ions is conserved and, hence, it does not depend on the electric potential.
The reason why we can consider the one-dimensional case as an approximation
is that protons are strongly
magnetized in the inner magnetosphere of Jupiter.
Hence, they are coupled to the magnetic field lines. The Larmour radius of
the thermal ions is of the order of $\rho_{Hi}^{(th)}\leq \mathrm{100\,m}$ and for energetic protons
it is $\rho_{Hi}^{(h)} \approx 2.5\,\mathrm{km}$. These scales are much
smaller than the size of the polar zone within which the electric potential can be
considered as a constant.

The flux of energetic electrons with the distribution function~(\ref{eq_4}) is
\begin{equation}
P_{e,z}^{(h)}=\frac{2\eta_e N_0 V_{e,h}}{\pi (1-\frac{e\varphi}{W_{e,h}})},
\label{eq_13}
\end{equation}
where $V_{e,h}=\sqrt{\frac{2W_{e,h}}{m}}$.
To determine the coefficient $\eta_e$ we need to equate the model
flux of energetic electrons within some energy
channel $\Delta P_{e,z}^{(h)}$ far away from the surface with the measured one
\begin{equation}
\Delta P_{e,z}^{(h)}=\frac{6}{\pi}\eta_e N_0w_{e,h}^{3/2}\int_{w_1}^{w_2}\frac{x^2}{(x+w_{e,h})^4}\,dx,
\label{eq_14}
\end{equation}
where $w_1=v_{min}^2$, $w_2=v_{max}^2$ determine the energy channel
within which the flux of energetic electrons is measured,
$w_{e,h}=2W_{e,h}/m$.

The flux of energetic ions with the distribution function~(\ref{eq_6}) is easily
calculated:
\begin{equation}
P_{i,z}^{(h)}=\frac{\eta_i N_0 V_{i,h}}{48\pi}. \label{eq_15}
\end{equation}
The parameter $\eta_i$ that enters eq.~(\ref{eq_15}) can be determined in the same way
as for electrons.

The fluxes of the secondary electrons from the surface of the moon caused by
thermal and energetic electrons are obtained with the help of eqs.~(\ref{eq_8}) and (\ref{eq_9}).
The yield $\delta (w)$ has its maximum at rather moderate energies $W_m$, which are much
smaller than the typical energy of energetic electrons. Also we suppose (it will be confirmed by calculations) that
the electric potential on the surface exceeds significantly the  energy
$W_m$, that is $|e\varphi_s| >>W_m$.  In this case the flux of the
secondary electrons produced by energetic electrons takes the form
\begin{equation}
P_{e,z}^{(sec,h)}\approx\frac{4N_{e,h}V_{e,h}^3v_{s,z}}{\sqrt{\frac{-2e\varphi_s}{m}}(w_{e,h}-\frac{2e\varphi_s}{m})^4}
\int_0^{\infty}\delta (w)w^2\,dw .\label{eq_16}
\end{equation}
The flux of the secondary electrons that correspond to the thermal primary
electrons takes the form
\begin{equation}
P_{e.z}^{(sec,th)}\approx 9.1\delta_m v_{s,z}N_0\frac{W_{e,th}}{W_m}\exp\left (\frac{e\varphi_s}{T_{th}}\right ),
\label{eq_17}
\end{equation}
where $v_{s,z}$ is the z-component of the velocity of the secondary electrons ($v_{s,z}\approx \frac{2}{\pi} v_s$
for electrons  with equal probability emitted from the surface in different directions). Suppose that the surfaces
of both moons are covered with  regolith which has very small electric conductivity
(it can be considered almost an insulator). In this case the stationary electric potential on the
surface is determined by the local relation
\begin{equation}
P_{e,z}^{(th)}(\varphi_s)+P_{e,z}^{(h)}(\varphi_s)-P_{e,z}^{(sec)}(\varphi_s)=P_{i,z}^{(th)}+P_{i,z}^{(h)}.
\label{eq_18}
\end{equation}
The flux of electrons produced by the solar UV radiation is absent in eq (18). As it was mentioned in
the Introduction
the role of this radiation in charging of the polar regions is negligible even
on the sunlit side.
After the substitution of the fluxes of electrons and ions presented above we are in a position to find
from eq.~(\ref{eq_18}) the surface electric potential. This will be done in the next section.

\section{Estimates of the surface electric  potentials for  Thebe and Amalthea}

In order to find the value of the electric potential on the surface from eq.~(\ref{eq_18}) we
need to provide concrete values for the input parameters. According to \citet[][see their Fig. 10]{divine1983},
the concentration of thermal plasma at the
orbit of Thebe is $N_0(3.1 R_J)\approx 50\,\mathrm{cm^{-3}}$, and at the orbit of Amalthea it is
$N_0(2.5 R_J)\approx 100\,\mathrm{cm^{-3}}$. The temperature of thermal plasma is
$T_e=T_i=46\,\mathrm{eV}$. Note that the main thermal ions in the inner magnetosphere are not protons
but heavy ions $O^{+}$ and $S^{+}$. Available experimental data show that the
strongest omnidirectional flux of energetic electrons
corresponds to the energy channel $0.19\,\mathrm{MeV} < W <0.26\,\mathrm{MeV}$ and for protons it
corresponds to the channel $1.2\,\mathrm{MeV} < W <2.15\,\mathrm{MeV}$.
For the orbit of Thebe the flux of electrons in a given channel is
$\Delta P_e^{(h)}\approx 6\cdot 10^8\,\mathrm{cm^{-2}s^{-1}}$, for the orbit of Amalthea it is an order of magnitude less
$\Delta P_e^{(h)}\approx 6\cdot 10^7\,\mathrm{cm^{-2}s^{-1}}$. The flux of ions in a given channel
for the orbit of Amalthea is $\Delta P_i^{(h)}\approx 6\cdot 10^6\,\mathrm{cm^{-2}s^{-1}}$, which is slightly less than the value for Thebe:
$\Delta P_i^{(h)}\approx 10^7\,\mathrm{cm^{-2}s^{-1}}$.
We shall use these fluxes to determine the coefficients $\eta_e$ and $\eta_i$. The
parameters for secondary electrons (maximum yield $\delta_m$ and the
corresponding energy $W_m$) vary significantly depending on the surface
material. The energy $W_m$ can be of the order of approximately $100 -
1000\,\mathrm{eV}$ while the maximum yield $\delta_m$ varies from less than unity up to 30.
In our calculations we take $W_m=200\,\mathrm{eV}$ and several different values
of $\delta_m$.

We shall use the model loss-cone distribution~(\ref{eq_4}) for energetic electrons in the vicinity of Thebe and  Amalthea.
Comparison  with the
experimental data  makes it possible to estimate the fraction of
energetic electrons~(\ref{eq_4})  $\eta_e\approx 4\cdot 10^{-3}$ and ions~(\ref{eq_6}) $\eta_i \approx 0.04$ for Thebe's
orbit. After the substitution of fluxes eq.~(\ref{eq_18}) takes the form
\begin{equation}
(12-5.3\delta_m )\exp\left (\frac{e\varphi_s}{T_{th}}\right )+\frac{2.8}{1-\frac{e\varphi_s}{W_{e.h}}}
-\frac{4\cdot 10^{-5}\delta_m}{\left (-\frac{e\varphi_s}{W_{e,h}}\right )^{1/2}\left (1-\frac{e\varphi_s}{W_{e,h}}\right )^4}=
0.1(1+\kappa) . \label{eq_19}
\end{equation}
Here $\kappa=\kappa (\varphi_s)$ is introduced to qualitatively take into account  a possible correction of
the 1-D model.
In general $\kappa$ depends on the surface potential $\varphi_s$. In different situations
its value can vary significantly. In the 1-D case (for very strong magnetic field when ions are attached
to the magnetic field lines) $\kappa =0$. If the Larmour radius of ions is much larger
than an object (e.g. dust grain), the 1-D approximation fails.
The physical cross-section of a negatively charged dust grain is larger than its geometric
cross-section due to the attraction of
positively charged ions, and $\kappa$ tends to a well-known value
$\kappa \approx -\frac{e\varphi_s}{T_{th}}$. In a strong
magnetic field the transverse drift of ions (which causes the deviation from the 1-D model) should be small enough.
In this case we expect only a small deviation from the 1-D model.
The first term in eq.~(\ref{eq_19}) is the difference between the fluxes of thermal
primary electrons and the secondary electrons generated by these primaries.
The next term describes the flux of energetic electrons. The last term on the
left-hand side corresponds to the contribution of secondary electrons
generated by energetic electrons from the radiation belt. The right-hand side
gives the input of thermal and energetic ions. Analysis of this equation
shows that the surface potential is determined by energetic electrons and not
by thermal electrons while $0.1(1+\kappa)<2.8$. The role of the secondary electrons produced by
energetic electrons hitting the surface is
negligible compared to all other fluxes that determine the surface potential, see eq. (19). Nevertheless the flux of these electrons at the surface is finite and it still plays a significant role in high charging of some dust grains on the surface of the moon \citep[the mechanism suggested by][]{wang2016}. The absolute value of this potential
\begin{equation}
|e\varphi_s|\approx\left (\frac{28}{1+\kappa}-1\right ) W_{e,h} \label{eq_20}
\end{equation}
is very high, e.g. $e\varphi_s \approx -2.1$ MeV, if we take $\kappa =1$. Such a high value of the surface potential
 is determined first of all by the magnitude
of the flux of energetic electrons. The stronger this
flux the higher is the absolute value of the surface potential. This is in accordance with the laboratory
experiments by \citet{wang2016} where it is shown that "when the energy of
electrons was increased gradually to 120 eV, the surface potential simply
followed to be increasingly negative, reaching $\sim 120\,\mathrm{V}$...".
A somewhat similar situation occurs
in the shadow of the Earth Moon when solar energetic particles (SEP) reach the vicinity of our
planet. In the SEP event the estimated potential of the Moon's surface in the
shadow grows many times (up to -4000 V) compared to the usual situation, see
e.g. \citet{halekas2007}. The second factor that influences the magnitude of
the surface potential is the distribution function of the energetic electrons.
 To demonstrate this we introduce another model loss-cone distribution which
decreases stronger for high energies of electrons than the first loss-cone distribution given by eq.~(\ref{eq_4})
\begin{equation}
F_e^{(h)}=\frac{48}{\pi^2}N_{e,h}\frac{v_{\perp}^2V_{e,h}^5}{(v_z^2+v_{\perp}^2+V_{e,h}^2)^5}. \label{eq_21}
\end{equation}
After the substitution of the corresponding flux of energetic electrons into eq.~(\ref{eq_19}) we
find the electric potential on the surface of Thebe:
\begin{equation}
|e\varphi_s|\approx\left (\sqrt{\frac{53}{1+\kappa}}-1\right ) W_{e,h}. \label{eq_22}
\end{equation}
The obtained potential is smaller than the previous one given by eq.~(\ref{eq_20}) (approximately 2.5 times for $\kappa =1$),
that is $e\varphi_s \approx -800$ KeV.
Note that the suggested two loss-cone distributions decrease with the growth of the electron energy somewhat
similar to the kappa distribution introduced earlier \citep{divine1983}.
The equation describing the surface potential at the polar region of Amalthea
is slightly different because the flux of energetic electrons in the vicinity
of Amalthea is smaller. The corresponding  equation takes the form:
\begin{equation}
(12-5.3\delta_m)\exp\left (\frac{e\varphi_s}{T_{th}}\right )+\frac{0.2}{1-\frac{e\varphi_s}{W_{e,h}}}=0.09(1+\kappa ).
\label{eq_23}
\end{equation}
In eq.~(\ref{eq_23}) we have neglected a small contribution of secondary electrons produced
by energetic primaries. The approximate solution of
eq.~(\ref{eq_23}) for small enough $\kappa \leq 1$ is:
\begin{equation}
|e\varphi_s|\approx\left ( \frac{2.2}{1+\kappa}-1\right )W_{e,h}. \label{eq_24}
\end{equation}
The solution of eq.~(\ref{eq_24})  corresponds to a smaller
surface electric potential compared to eq.~(\ref{eq_20}). Taking  $\kappa
=0.7$ we obtain the surface potential $e\varphi_s =-49$ KeV.
It should be mentioned that with the growth of the flux of ions
the absolute value of the surface potential rapidly decreases and
the contribution of thermal electrons to the total flux increases. The fluxes of thermal
and energetic electrons become equal at $e\varphi_s\approx -161\,\mathrm{eV}$ (for $\delta_m =1, \kappa\approx 3$).
If the flux of ions continues to grow, thermal electrons
begin to determine the electric potential on the surface:
\begin{equation}
e\varphi_s \approx T_{th} \ln\left (\frac{0.09(1+\kappa)-0.2}{12-5.3\delta_m}\right ).
\label{eq_25}
\end{equation}
Such a transition from a very strong surface potential (eq.~\ref{eq_24}) to a smaller one (eq.~\ref{eq_25}) is
more probable in case of Amalthea than in case of Thebe.

\section{Electric fields above the surfaces of Thebe and Amalthea}

The aim of this section is to give analytical estimates of the electric
fields  above Thebe and Amalthea in their polar regions. The procedure is
straightforward. We need to substitute concentrations of electrons and ions
(found from their distribution functions) into the Poisson equation:
\begin{equation}
\frac{d^2 \varphi}{dz^2}=4\pi e\left (N_i^{(th)}(\varphi ) + N_i^{(h)}(\varphi )-N_e^{(th)}(\varphi )-
N_e^{(h)}(\varphi )-N_e^{(sec)}(\varphi )\right ). \label{eq_26}
\end{equation}
Here, $N_i^{(th)}$ and $N_i^{(h)}$ are the concentrations of thermal and
energetic ions above the surface at some height $z$, $N_e^{(th)}$ is the concentration of
thermal electrons at the height $z$ which consists of electrons reaching the surface and which are reflected
by a strong electric potential below a given height, $N_e^{(h)}$ is the
concentration of  energetic electrons, and $N_e^{(sec)}$ is the concentration of secondary electrons created
at the surface of the moon. Eq.~(\ref{eq_26}) should be solved  with the boundary conditions at the
surface $\varphi (z=0)=\varphi_s$ and very far above the surface $\varphi
(z=\infty )=0$. Unfortunately, even in the 1-D case this equation can be solved
only numerically. Nevertheless, it is possible to give reliable estimates of
the distribution of the electric potential above the surface in analytical
form.

Again we assume that the deviation from  the 1-D case in the motion of ions just above the surface is small.
In this case the electric potential in the polar region of Thebe
is given by eq.~(\ref{eq_20}), and for Amalthea by eq.~(\ref{eq_24}).
Note that far above the surface the quasi-neutrality in the 1-D case cannot be supported.
Indeed, all ions approaching from above along the field line are absorbed by the surface while almost all
thermal electrons are reflected by a strong negative potential. At the same time ions coming from below in the 1-D case
are screened by the moon and hence cannot contribute to quasi-neutrality. In order to obtain
 quasi-neutrality of the plasma
we need to consider that the magnetosphere of Jupiter moves with the
rotation of Jupiter which is faster than
the orbital velocity of Thebe or Amalthea. Due to this, ions coming from below along the magnetic field lines
contribute to the restoration of quasi-neutrality in the plasma far above the moon's surface.

We would like to find analytically the vertical electric field above the surface. For this aim we shall use a
slightly modified  procedure suggested by \citet{borisov2006}. Estimates
show that for  strong surface potentials the main contribution to the left-hand side of the Poisson equation~(\ref{eq_26})
just above the surface gives the concentration of thermal ions. Therefore, we proceed with the following
approximate equation:
\begin{equation}
\frac{d^2 F}{d z^2}\approx\frac{1}{2R_D^2}\frac{1}{\sqrt{\pi F(z)}}. \label{eq_27}
\end{equation}
Here $F=-\frac{e\varphi}{T_{th}}$, $R_D=V_{e,th}/\omega_{Pe}$ is the Debye radius,
$\omega_{Pe}=\left (\frac{4\pi e^2N_0}{m}\right )^{1/2}$ is the plasma frequency of thermal electrons.
Multiplying both sides with $dF/dz$  and integrating with respect to $z$ we find:
\begin{equation}
\frac{dF}{dz}=-\frac{1}{R_D}\left (\frac{2}{\sqrt\pi}(C_0+\sqrt F)\right )^{1/2},
\label{eq_28}
\end{equation}
where $C_0$ is an unknown constant. Assuming that the absolute value of $C_0$
is much less than the dimentionless potential $F_s=-\frac{e\varphi_s}{T_{th}}$
on the surface of the moon \citep[for details see][]{borisov2006}, we find an approximate value
for the potential $F$ above the surface in the polar region:
\begin{equation}
F\approx F_s-\frac{\sqrt 2}{\pi^{1/4}}F_s^{1/4}\frac{z}{R_D}.
\label{eq 29}
\end{equation}
Potentials $F_s$ for Thebe and Amalthea can be found from eq.~(\ref{eq_20}) and eq.~(\ref{eq_24})
respectively. Assuming as an
example that for Thebe $\kappa =1$, we find that $e\varphi_s\approx - 2.1$ MeV. Now we
can estimate the electric field just above the surface of
Thebe:
\begin{equation}
\frac{d\varphi}{dz}\approx \frac{T_{th}}{eR_D}\frac{2^{1/2}}{{\pi}^{1/4}}F_s^{1/4}.
\label{eq_30}
\end{equation}
It follows from eq.~(\ref{eq_30}) that the electric field $E_z(0)$ depends on the surface potential
weakly enough $\propto F_s^{1/4}$. For Thebe if the surface potential is $e\varphi_s\approx - 2.1$
MeV, the electric field is estimated as $E_s\approx -0.71\,\mathrm{V\,cm^{-1}}$.

Similarly, we can find the electric field at the surface of Amalthea. Assuming for example that
$\kappa =0.7$ we find from eq.~(\ref{eq_24}) that the surface potential for Amalthea
 is $e\varphi_s\approx -49$ KeV and the surface electric field
is $E_s\approx -0.39\,\mathrm{V\,cm^{-1}}$.
Thus, even though the negative potential on the surface is very strong (especially for Thebe),
the electric field is quite moderate. The electric field slowly decreases with the height above the
surface according to
eq.~(\ref{eq_28}). This decrease can be expressed in an explicit form:
\begin{equation}
|eE_z|\approx \frac{T_{th}}{R_D}F_s^{1/4}\left
(1-\frac{0.25}{F_s^{3/4}}\frac{z}{R_D}\right ). \label{eq 31}
\end{equation}
It follows from eq. (31) that this decrease at some distance above
the surface (while the constant $C_0$ in eq.~(\ref{eq_28}) can be neglected) is very weak. The
characteristic scale is of the order of $\Delta z \sim R_D F_s^{3/4}$.
With the help of these estimates
lofting of charged dust grains above the surface of Thebe  and Amalthea
will be discussed in Section~7.

\section{Action of adhesion force and multiple electric charges on dust grains}

It is well-known that the adhesion force and gravity can prevent the lofting of
dust grains from the surface of a cosmic body.
The magnitude of the adhesion force  between dust grains lying on
the surface as a function of size
was discussed in many publications, see, e.g.
\citet{hartzell2011,hartzell2013a,hartzell2013b,kimura2014,cooper2001} and  references therein. According to
the theory  the adhesion force between dust grains depends on their contact area
and material of grains (also the surface chemistry is important).
Theoretical estimates show that the adhesion force is large and to overcome it, strong
electric fields are required.
For example, for micron size grains according to \citet{hartzell2011,kimura2014}
the electric field should exceed $10^2-10^3~V\,\mathrm{cm^{-1}}$ and even more. 
Note that for irregular grains
according to \citet{cooper2001} the adhesion force varies by several orders of magnitude. Thus,
significant uncertainty in the magnitude of this force should be taken into account.

Several different ideas were introduced to explain how such electric fields
can be formed at the surface of a cosmic body. It was argued by \citet{criswell1973,rennilson1974,de1977}
that near the terminator on a flat surface strong electric fields can appear
between two points if one of them is exposed to the solar UV radiation and
the other point is in shadow. Later on it was shown theoretically
that at the terminator and in the shadow the roughness of the surface can help
the formation of strong electric fields. Indeed, according to \citet{borisov2006}
non-equal charging of two different slopes of a small cavity (or a hill) by the solar wind electrons and
protons can produce significant local electric fields. Recently it was
demonstrated in a laboratory experiment by \citet{wang2016} that in the presence of thermal plasma the action of
the electron beam on a
rough porous surface can create very strong electric charges
on some dust grains  (much larger than one elementary charge) due to the secondary electron emission.
In such a case sub-micron and micron size grains begin to move on the surface. According to estimates presented by
\citet{wang2016,schwan2017}  dust grains lofted from the surface have electric charges orders
of magnitude higher than the elementary charge. This experimental result, despite
some constraints in the explanation (positive electric charges should be taken into account),
provides the most convincing argument that charged dust grains
lying on the surface can overcome the adhesion force.

Therefore, due to bombardment by energetic charged particles,
two types of electric fields  can be singled out in the vicinity of the rough surface of a porous cosmic
body.
One of them is strong local electric fields that appear because energetic
electrons and ions penetrating into the body produce significant additional
ionization. This process results in positive charging of ionized atoms and
emission of secondary electrons. Note that positive and negative electric charges are
distributed stochastically on dust grains in the surface layer.
According to the experiment mentioned above the electric
force acting on some of the grains possibly exceeds the adhesion force and liberates such grains. The second
electric field is a large-scale one acting in the plasma sheath above the
surface. This regular electric field is much smaller than the typical
local field and can be estimated by integration of the Maxwell equation
$\nabla {\bf E}=4\pi\rho_e $ ($\rho_e$ is the electric density) across the
layer where electric charges are concentrated: $E_z(+0)-E_z(-0)=4\pi\int \rho_e(z_1)\,dz_1,$ where $E_z(+0)$ is the
vertical component of the electric field just above the surface and $E_z(-0)$
is the vertical component of the electric field below the surface layer. Note
that the amount of positive electric charges that appear in the surface layer of the cosmic body due to ionization
of atoms should be equal to the negative charges of the liberated secondary electrons
(assuming that all these electrons remain inside the cosmic body attached to dust
grains). In such a case the electric field in the plasma sheath above the surface $E_z(+0)$ is
connected with the excess of negative charges on the surface produced by the
fluxes of charged particles hitting the surface. Only this field determines the
dynamics of charged dust particles above the surface. In the next section we will discuss
the motion of charged dust grains
in the plasma sheath under the action of the electric forces and  gravity.

\section{Lofting of dust grains in the polar regions of Thebe and Amalthea}

As it was mentioned in the Introduction the lofting of charged dust grains in the electric fields above
the surface of the Moon and asteroids
has been discussed in many publications. As for Jupiter's moons, so far
only one mechanism (micrometeoroid impacts) was considered as a source of dust above the
surface. In this section we shall investigate the lofting in the polar regions of
Thebe and Amalthea of charged
dust grains that are able to overcome the action of
the particle adhesion. Our aim is to check if the lofting of charged dust grains in
the polar regions of Thebe and Amalthea can exceed the escape velocities in the electric fields.

The equations describing the lofting of charged dust grains above the
surface are:
\begin{eqnarray}
M_d\frac{d^2 z}{dt^2} & = & Q_d(t) E_z(z) -g_m(z)M_d \nonumber \\
\frac{dQ_d}{dt}       & = & e \sigma_d [P_i(\varphi (z))-P_e(\varphi (z))+P_{uv}]. \label{eq_32}
\end{eqnarray}
Here $M_d\approx 4\pi\rho_d a^3/3$ is the mass of a dust grain, $a$ and $\rho_d$
are its radius and its density, $g_m(z)$ is the gravity, $g_m(z)=g_m(0)\frac{R_m^2}{z^2+R_m^2}$,
$g_m(0)$ is the gravity on the surface of the moon, $R_m$ is the radius of the moon, $E_z(z)$ is
the electric field above the surface that changes according to eq. (31), $Q_d$ is the electric charge on a given dust
grain, $\sigma_d$ is the cross-section of a dust grain, $P_e, P_i$  represent the total
fluxes of electrons and ions acting on a dust grain with some potential $\varphi$ at the height $z$.
In the general case they are determined
by the distributions of  different components given in Section 2. The last term in eq. (32)
 $P_{uv}$ is the flux of photoelectrons
produced by the action of the solar UV radiation on the grain at the sunlit
side \citep{horanyi1991}:
$P_{uv}= ef$, if $\varphi <0$, and $P_{uv}=ef \exp(-e\varphi/T_{ph})$, if $\varphi
>0$. Here $f\approx 2.5*10^9\,\mathrm{cm^{-2}s^{-1}}$ for dielectrics, $T_{ph}\approx
(1-3)$ eV is the temperature of the emitted photoelectrons.
The boundary conditions
for eqs.~(\ref{eq_32}) are:
\begin{equation}
z(0)=0,~~dz/dt|_0=0,~~Q_d(0)=Q_0. \label{eq_33}
\end{equation}
Here $Q_0$ is the electric charge on a dust grain lying on the surface.
Note that our aim is to discuss only the initial stage of the motion of dust grains.
Unfortunately, the gravities of Thebe and Amalthea are not well-known. It is accepted that for Thebe
the surface gravity is $g_m\approx 1.3\,\mathrm{cm\,s^{-2}}$ and for Amalthea $g_m\approx 2\,\mathrm{cm\,s^{-2}}$.
At the same time there are some indications that Amalthea (possibly Thebe  also) is rather porous,
see \citet{anderson2005}. In this case
their gravity could be smaller by a factor of two ($g_m\approx 0.65\,\mathrm{cm\,s^{-2}}$
for Thebe and $g_m\approx 1\,\mathrm{cm\,s^{-2}}$ for Amalthea).

As it was shown in the previous section, the averaged electric density of the charged surface
$\Sigma_e=\int \rho_e(z)\,dz$ is connected with the vertical electric field.
If the surface is flat we find from this relation and equation $E_z(+0)-E_z(-0)=4\pi\Sigma_e,$
that on a given dust grain with the radius $a$ the electric charge is $q\approx 0.5 E_z(+0) a^2$.
 For the vertical electric fields above Thebe ($E_z=-0.71\,\mathrm{V\,cm^{-1}}$) and Amalthea
($E_z=-0.39\,\mathrm{V\,cm^{-1}}$)
dust grains with the radius $a\leq 10\,\mathrm{\mu m}$ lying on the surface should have at maximum one
elementary charge (i.e. either one electron or nothing).
Note that in reality as shown in laboratory experiments \citep{wang2016,schwan2017} for a porous surface layer
the electric charge on a given dust grain can be 2-3 orders of magnitude higher than estimated
above. These experiments clearly demonstrate that some highly charged
dust grains lying on the surface overcome the short-range adhesion force. The typical velocities
above the surface in the
experiment by \citet{wang2016} were approximately $0.5\,\mathrm{m\,s^{-1}}$. We are interested in much
higher velocities (of the order of the escape velocities). That is why the
initial velocity in eqs. (33) is assumed to be zero.

After lofting from the surface, a dust grain with electric charge $Q_0$ continues to acquire
a negative electric charge in the shadow because the flux of electrons
above the surface exceeds the flux of ions. At the initial stage the potential
on the charged grain is much smaller than the potential in space. Hence, the fluxes of electrons
and ions to the surface
of a grain are determined by the local electric potential in space $\varphi$ given by eq.~(\ref{eq_26}).
The absolute value of the electric potential on a grain $\varphi_d=Q_d(t)/a$
grows in time until the flux of electrons on its surface becomes equal to the flux of
ions. At the same time estimating in the shadow the flux of electrons that deposit their charge
on a dust grain lofting from the surface,
we need to take into account that not all electrons
 can be stopped by a small dust grain. For a dust grain with
radius $a\sim 0.5~\mu\mathrm{ m}$ we need to take into account the fluxes of electrons with
the energies $W\leq 1\,\mathrm{keV}$.
Near the surface with very strong negative potential the flux of such electrons is weak. Therefore,
the electric charge on a given dust grain above the surface  grows very slowly  (the characteristic time
can be of the order of an hour and even more, depending on the size of a grain).

 First, we estimate the radius of a  grain with a
single charge $e$ that can overcome the surface gravity $g_m$:
\begin{equation}
a < a_{cr}=\left (\frac{3eE_z(0)}{4\pi \rho_d g_m}\right )^{1/3}, \label{eq_34}
\end{equation}
where $\rho_d$ is the density of dust which we  take as $\rho_d\approx
1\,\mathrm{g\,cm^{-3}}$. Substituting in eq.~(\ref{eq_34}) the estimates obtained above  for the
surface electric field, we find that for Thebe $a_{cr}=0.47~\mu\mathrm{m}$ or $a_{cr}=0.59~\mu\mathrm{m}$,
respectively, depending on the surface gravity. Similarly, for Amalthea
$a_{cr}=0.418~\mu\mathrm{m}$ or $a_{cr}=0.52~\mu\mathrm{m}$. As we can see from these estimates the difference
between critical radii due to the uncertainty of the gravity is negligible. At the same time,
as shown in the laboratory experiment \citep{wang2016,schwan2017}
the electric charge on a given dust grain can be much higher than a single elementary charge. In this case the critical size
for lofting grows significantly. For example, if the electric charge on a dust grain
is $Q=10^2~ e$ the critical sizes exceed several times the values given above.

Our aim is to estimate the dynamics of a given grain above the surface.
It should be mentioned that the charged dust grain moves not only vertically but also in the horizontal plane.
Due to this, after some time the grain  leaves the polar region where a strong negative potential exists.
Hence, to discuss the trajectories of dust grains for longer time intervals we have to add
the system of equations describing the
dynamics of a charged dust grain
subjected to $[{\bf V}\times {\bf H}]$ electric and
magnetic fields, the gravity of Jupiter, and the solar UV radiation \citep{horanyi1991}. This will be
done in a separate paper.

Before proceeding with the numerical calculations we would like to present
analytical estimates of the sizes of charged dust grains that can leave the moon
(Thebe or Amalthea) in electric fields. We shall take for the escape velocity
of Thebe $V_{esc}=25\,\mathrm{m\,s^{-1}}$ and the density $\mathrm{\rho_d =1\,g\,cm^{-3}}$. It follows from the conservation of energy
that a dust grain with the charge $Q=100~e$ and the
radius $a$ lofting from the surface with the  potential $e\varphi_s=-2.1$ MeV  leaves the moon if $a < 2.2~\mu m$. For
smaller potential $e\varphi_s=-800\,\mathrm{KeV}$ the size should not exceed
$a=1.6~\mu m$. For Amalthea assuming the surface potential to be $e\varphi_s
=-49$ KeV, see the end of Section~5, and the escape velocity $V_{esc}=\mathrm{58\,m\,s^{-1}}$ we find that
to leave the moon, dust grains should have a size $a< 0.5~\mu m$. These
estimates correspond to  sizes of grains detected above Thebe and
Amalthea, respectively.

Now we present the results of numerical calculations for the initial stage of dust grains lofting from
Thebe and Amalthea in the shadow. We assume that some dust grain with  radius
$a$ lying on a rough surface in the polar region of Thebe (or
Amalthea) has a negative charge 2-3 orders of magnitude higher than in the case of a
flat surface \citep{wang2016,schwan2017}.
In our calculations we take for the
surface gravity of Thebe the value $g_m(0)=1.3\,\mathrm{cm\,s^{-2}}$ and for
Amalthea $g_m(0)=2\,\mathrm{cm\,s^{-2}}$. The calculations are carried out for
periods  when changes in the electric charge on a given
dust grain can be neglected. The averaged radius of Thebe is $R_m=49$ km, while that of
Amalthea is $R_m=83.4$ km, the Debye radius $R_D$
for Thebe is $R_D\approx 10$ m, while for Amalthea it is $R_D \approx 7.14$ m.
In Figure~1 we present the velocities of two dust grains with radii
$a=0.4~\mu\mathrm{m}$ (continuous line)
and $a=1~\mu\mathrm{m}$ (broken line)  as functions of time. The negative electric charge on each
grain is taken as $Q_0=100\,e$. The electric field on the surface of Thebe in
accordance with our estimates is $E_z(0)=0.71\,\mathrm{V\,cm^{-1}}$. The horizontal line is
the escape velocity for Thebe taken as $V_{esc}=25\,\mathrm{m\,s^{-1}}$. The vertical
electric field slowly decreases with the height, see eq. (31). The calculations are terminated when
both grains acquire the escape
velocity. Figure~2 demonstrates the growth of velocities of the same dust grains above
Amalthea's surface. The electric field on the surface of Amalthea is taken as
$E_z(0)=0.39\,\mathrm{V\,cm^{-1}}$. The horizontal line shows the escape velocity for
Amalthea which is taken as $V_{esc}=58\,\mathrm{m\,s^{-1}}$.
It is seen that a small grain above the polar region of Thebe achieves the escape velocity very
quickly (in 7 sec), while for a larger grain it requires much more time (100~sec).
The smaller grain above Amalthea exceeds the escape velocity in 25~sec.
At the same time the larger grain does not exceed the escape velocity at all. Note
that from these figures we can estimate the height above the surface at which the escape velocity is achieved.
For smaller grains ($a=0.4~\mu\mathrm{m}$) the corresponding height above Thebe is $z_{esc}^{(Th)}\approx 87$ m
while for Amalthea it is $z_{esc}^{(A)}\approx 750$ m.

\section{Discussion and conclusions}

We have discussed the electric charging of the polar regions of Jupiter's moons Thebe and
Amalthea and dust grains lofting from them. According to our analysis
the lofting of charged dust grains in electric fields from the surfaces of the moons 
could contribute to the dust population in the gossamer rings, in addition to the main mechanism - secondary ejecta
released by impacts of micrometeorites onto the surfaces. The suggested
mechanism can explain some peculiarities of the dust
distribution within the gossamer rings. First of all we argued that the height distribution
of dust in the vicinity of Amalthea and especially for Thebe  should be  broader and smoother than it
was modelled before. Micrometer- and sub-micrometer-sized dust grains in the vicinity
of Thebe are accelerated in the electric field
of the double layer and acquire a velocity sufficient to overcome the gravity of the moon.
Note that in the vicinity of the moons (especially for Thebe) a rather smooth
height distribution of dust was indeed detected, see e.g. Figure~13 in  \citet{showalter2008}.
Thus, our results qualitatively agree with the experimental data.

According to our analysis, the polar regions of Thebe and Amalthea  should have high negative potentials
due to bombardment by energetic electrons. This theoretical result is based on two main assumptions.
First, the electric conductivity of the moons' surfaces is assumed to be close to zero (pure insulator).
Second, we expect that
for the distribution  of the electric potential the 1-D approximation can be used. In reality the potential
significantly varies not only along the magnetic field line but also in
transverse directions. But the typical velocities of the energetic and even thermal ions near
the surface of the moons due to acceleration in the electric field are orders
of magnitude higher than the transverse drift velocity in a strong magnetic
field. Due to this, ions move more or less along the magnetic field lines.
 Energetic electrons that provide
the main input to the charging of the surface also have longitudinal
velocities much higher than the transverse drift velocity. Thus, a 1-D approximation as a first step for
calculating the surface charging can be justified.
The formation of strong double layers above the polar regions
 is an important factor that should be taken into account. These rather broad structures where
the concentration of the thermal plasma is smaller than in the inner magnetosphere of Jupiter
possibly can be detected by  future space missions.

\section*{Acknowledgements}
This work was partially performed during repeated visits of N. Borisov at MPS.
NB is grateful to MPS for financial support during these visits.



\newpage
\section{Figures}

\begin{figure}[hb]
\vspace{-4cm}
		\includegraphics[width=\textwidth]{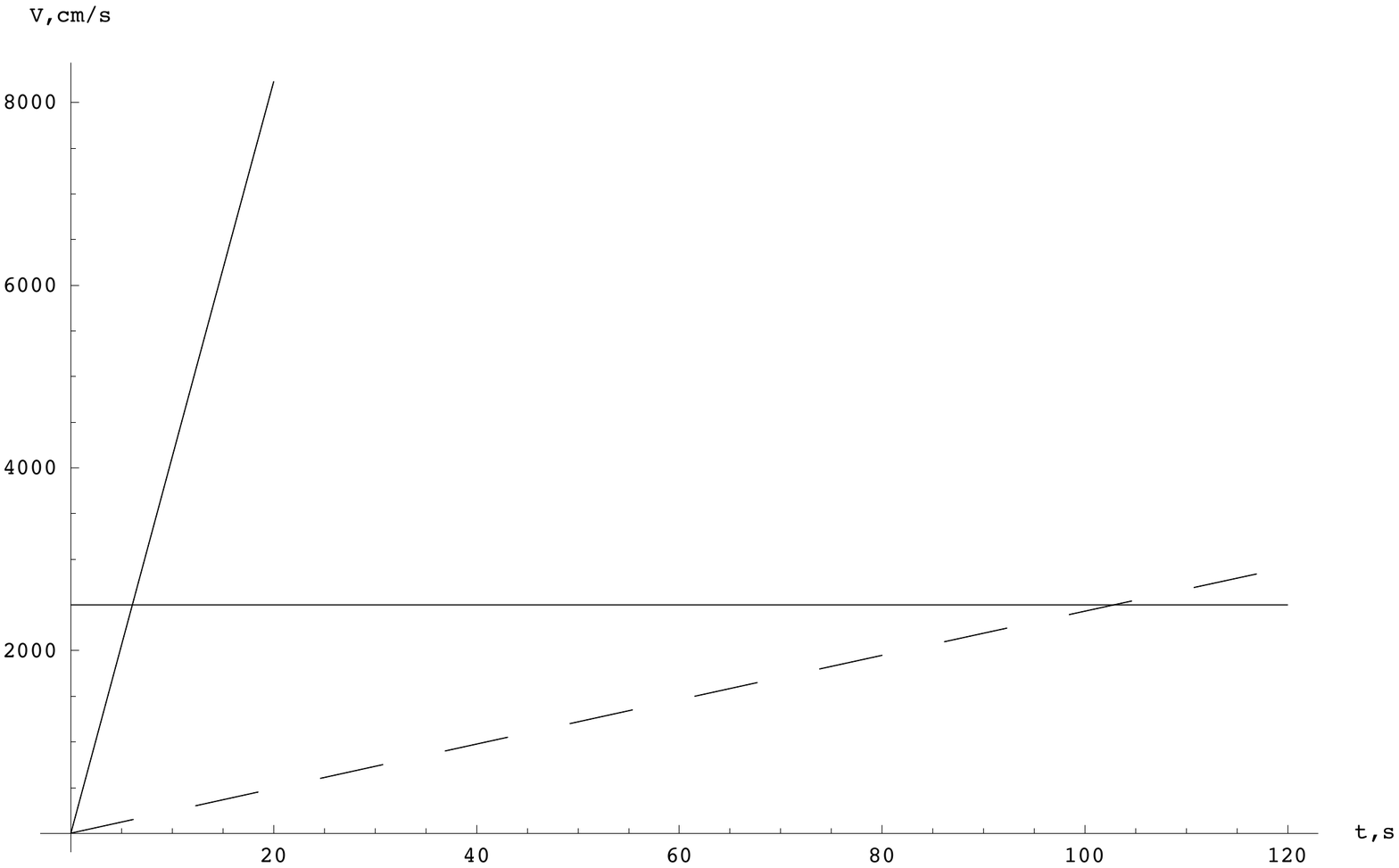}
		\vspace{-5cm}
	\caption{Vertical velocities of two dust grains with radii $a=0.4~\mu\mathrm{ m}$ (continuous line) and
$a=1~\mu\mathrm{ m}$ (broken line) in the shadow above Thebe as a function of time. The electric charge on
each grain is constant, equal to
$Q=100~e$. The gravity of Thebe is taken as $g_m\approx 1.3~\,\mathrm{cm\,s^{-2}}$.
The horizontal line $V=\mathrm{25\,m\,s^{-1}}$ corresponds to the escape velocity of Thebe.
		  }
	\label{fig:1}
\end{figure}

\begin{figure}[b]
\vspace{-6cm}
		\includegraphics[width=\textwidth]{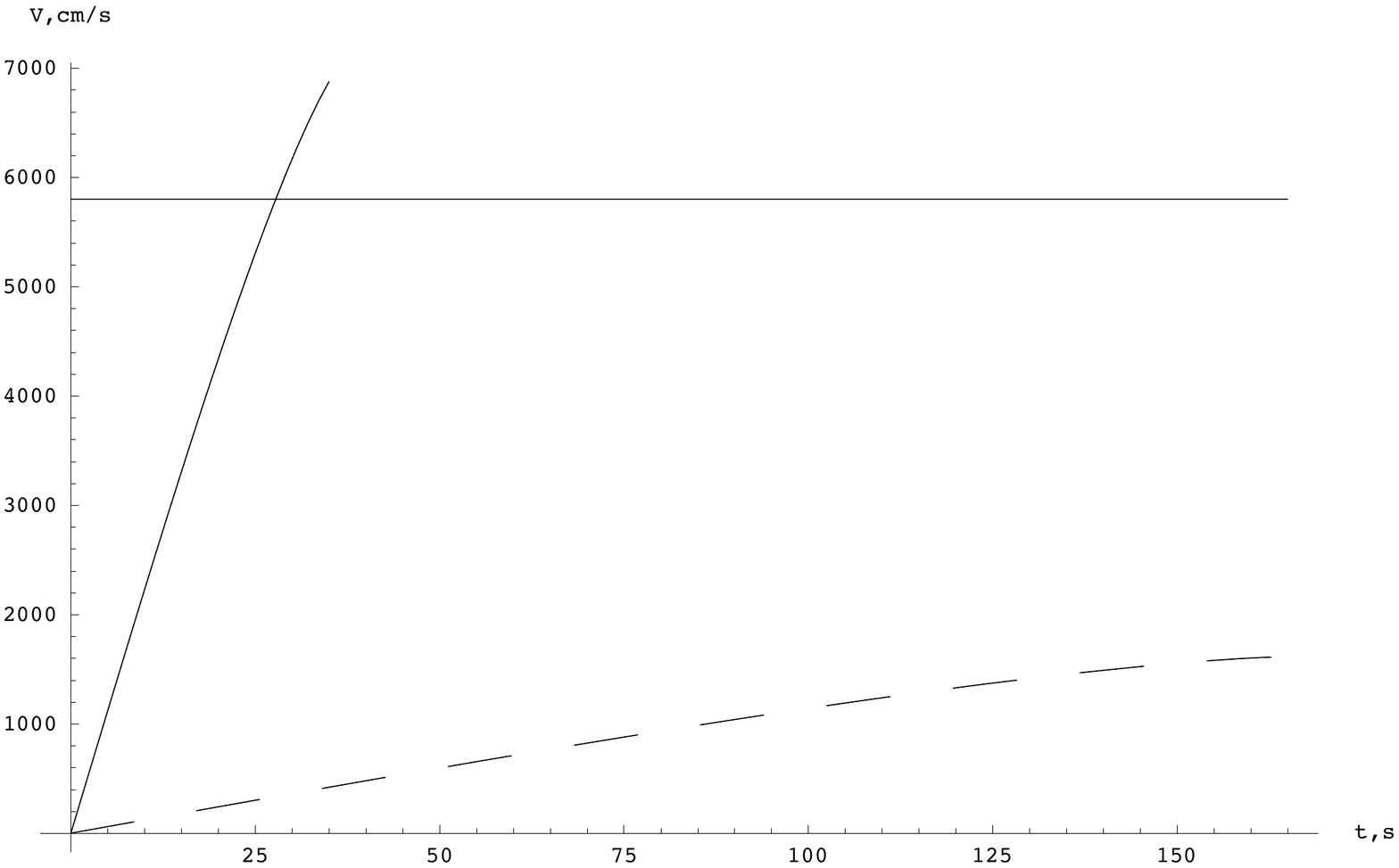}
		\vspace{-5cm}
	\caption{Vertical velocities of two dust grains with radii $a=0.4~\mu\mathrm{ m}$ (continuous line) and
$a=1~\mu\mathrm{ m}$ (broken line) in the shadow above Amalthea as a function of time. The electric charge
on each grain is constant, equal to
$Q=100~e$. The gravity of Amalthea is taken as $g_m\approx
2~\,\mathrm{cm\,s^{-2}}$. The horizontal line $V=\mathrm{58\,m\,s^{-1}}$ corresponds to the
escape velocity of Amalthea.
		  }
	\label{fig:1}
\end{figure}

\end{document}